\documentclass[12pt]{iopart}
\usepackage{iopams}
\usepackage{graphicx}
\begin{document}
\title[Non-universal quantities within functional renormalization group approach]
{Calculation of non-universal thermodynamic quantities within
self-consistent non-perturbative functional renormalization group
approach}

\author{V I Tokar$^1$}

\address{$^1$ Universit\'e de Strasbourg, CNRS, IPCMS, UMR 7504,
F-67000 Strasbourg, France}

\begin{abstract}
A self-consistent renormalization scheme suitable for the calculation of
non-universal quantities in $n$-vector models with pair spin interactions
of arbitrary extent has been suggested.  The method has been based
on the elimination of the fluctuating field components within the
layers defined by the layer-cake representation of the propagator. The
non-perturbative renormalization group (RG) equations has been solved in
the local potential approximation.  Critical temperatures of the $n=1-3$
vector spin models on cubic lattices have been calculated in excellent
agreement with the best known estimates. Several critical amplitudes and
the magnetisation curve of the Ising model on the simple cubic lattice
calculated within the approach compared well with the values from
literature sources. It has been argued that unification of the method
with cluster techniques would make possible the treatment of realistic
lattice models with multi-spin interactions and describe in a unified
framework the phase transitions of any kind. Besides the RG equations
the layer-cake technique can be used in the exact partial renormalization
of the local interactions in the functional lattice Hamiltonians.
This procedure reduces the strength of the interactions which in some
cases can make them amenable to perturbative treatment.
\end{abstract}

\submitto{\jpa}
\maketitle
\section{Introduction}
Phase transitions in many-body systems are formally defined as the
points where derivatives of the free energy with respect to thermodynamic
variables become singular. Microscopic Hamiltonians are usually assumed
to be smooth functions of their parameters, so calculations of the free
energy based on series expansion in the powers of the Hamiltonian or
of its parts to any finite order cannot exhibit the singularities. This
means that phase transitions can be described only within non-perturbative
approaches capable of calculating the free energy to all orders in the
Hamiltonian which in general is a very difficult task.

The problem simplifies in the case of lattice models where the infinite
system can be modelled by a finite cluster of sites.  The
partition function of the cluster can be calculated exactly to all orders
in the interaction parameters and with suitable measures taken to embed
the cluster into the infinite system remarkably accurate results can be
obtained with the use of relatively small clusters (see
\cite{ducastelle,maier_quantum_2005,tokar_new_1997,tan_topologically_2011}
and references to earlier literature therein).  The cluster approximation
is systematic in the sense that it can be indefinitely improved
by enlarging the cluster size so in principle it presents a viable
alternative to the series expansions as a general approach to the
many-body problems with strong interactions.

From the standpoint of the phase transitions theory the main drawback
of the cluster approach is its inability to adequately treat the
long-range fluctuations \cite{tokar_new_1997,tan_topologically_2011}
that are paramount to the description of the continuous
transitions and the critical points.  As is known, the critical
behaviour can be successfully described within the renormalization
group (RG) approach \cite{wilson}. Extensive studies in the
framework of the perturbation theory have revealed many aspects
of the universal behaviour in the vicinity of the critical point
\cite{wilson,le_guillou_zinn-justin,RG2002review}.  In particular,
accurate values of the critical exponents and the critical amplitude
ratios have been calculated which for many practical purposes can be
considered as exact and by virtue of the universality can be used in
interpretation of experimental data in a variety of systems. But for
a concrete system one is also interested in non-universal quantities
such as the absolute values of the critical amplitudes and the critical
temperature as well as in the behaviour beyond the critical region. A
natural way of achieving this would be the development of suitable
non-perturbative RG techniques in order to profit from the well developed
description of the universal behaviour. Besides, the RG approach is known
to be completely general and may be used also away from the critical
point \cite{wilson}.  The non-perturbative approximations to the exact RG
equations have been extensively studied over the past decades but existing
calculations of the system-specific non-universal quantities within this
approach did not lead to development of a general systematic technique
(a discussion of this problem and an extensive bibliography on the
non-perturbative RG may be found in \cite{berges_non-perturbative_2002}).

The aim of the present paper is to suggest a renormalization scheme
closely related to the functional cluster techniques introduced in
\cite{gamma_exp,tokar_new_1997}. The ultimate goal would be to develop a
non-perturbative RG approach that could be unified with the cluster method
in such a way as to make possible systematic improvement of the accuracy
thus making it also self-contained.  Besides, the unification would make
possible dealing with non-local many-body interactions which arise, e.g.,
in the realistic models of alloys \cite{ducastelle,Zunger2004}. However,
in the present paper only the RG equations will be considered in detail
because, on the one hand, the feasibility of such a unification discussed
in section \ref{discussion} is in principle rather straightforward,
but on the other hand, it would require extensive additional
calculations requiring a separate study \cite{arXiv16}. Besides,
the local potential approximation (LPA) that will be used in the
implementation of the RG scheme is not restricted to lattice models
\cite{wegner_renormalization_1973,nicoll_exact_1976,local_potential}
so the results obtained in the present paper may be straightforwardly
applied to the models in continuous space as well.
\section{Definitions and notation}
The partition functions of the lattice $n$-vector model can be represented
in the functional integral form as
\begin{equation}
	Z[{\bf h}]=\int D{\bf s} e^{-H[{\bf s}]+{\bf h}_i\cdot{\bf s}_i}
\label{Z}
\end{equation}
where the real fluctuating $n$-vector field ${\bf s}_i$ and the external
source field ${\bf h}_i$ are defined at the sites $i$ of a periodic
$d$-dimensional lattice of size $N$; $D{\bf s} =\prod_id{\bf s}_i$ and
$H$ is a dimensionless Hamiltonian. To simplify notation summation over
repeated {\em discrete} subscripts corresponding to the lattice sites
and the $n$-vector components will be implicitly assumed throughout the
paper. Boldface characters and the dot products will refer both to $d$-
and $n$-dimensional vectors, though for simplicity the site coordinates,
such as $i$, will not be boldfaced.

The connected correlation functions (CFs) of the field can be found by
differentiation of their generating functional $\ln Z[{\bf h}]$ with
respect to the source field ${\bf h}$ \cite{vasiliev1998}. Two CFs will
be calculated in the present paper : the magnetisation $m_{\sigma i}$
equal to the average
\begin{equation}
	\label{s_av}
	{m}_{\sigma i} \equiv \langle s_{\sigma i}\rangle
	= \left.\frac{\partial\ln 
		Z[{\bf h}]}{\partial h_{\sigma i}}\right|_{\bf h=0},
\end{equation}
where the subscript $\sigma=1,\dots,n$ numbers the vector components,
and the connected part of the pair CF $G^R_{ij}$ 
\begin{equation}
	\label{ss_av}
	G_{ij}^R= \left.\frac{\partial^2\ln Z[{\bf h}]}{\partial h_{\sigma i}
\partial h_{\sigma j}}\right|_{\bf h=0}
=\langle {s}_{\sigma i}{s}_{\sigma j}\rangle -
{m}_{\sigma i}{m}_{\sigma j}.
\end{equation}
Though in general non-trivial pair correlations between all field
components may exist, only the diagonal in $\sigma$ CF defined in
(\ref{ss_av}) will be used in the present study. Besides, because only
fully $O(n)$ symmetric Hamiltonians will be considered below and only
the symmetric phase treated in the $n>1$ case all diagonal CFs will be
equal and the subscript $\sigma$ may be omitted; the spontaneous symmetry
breaking will be discussed only for the $n=1$ Ising model in which case
the subscript is also superfluous.  The superscript $R$ in (\ref{ss_av})
and below will mark all fully renormalized quantities.

In the present paper we will deal with Hamiltonians of the following 
general form
\begin{equation}
H[{\bf s}]=\frac{1}{2}{s}_{\sigma i}\epsilon_{ij}{s}_{\sigma j}
+H_I[{\bf s}].
	\label{H}
\end{equation}
where the first part describes pair interactions between $n$-vectors at
different sites while the interaction part $H_I$ will be assumed to be
local to the sites; matrix $\hat{\epsilon}=||\epsilon_{ij}||$ does not
depend on $\sigma$ because of the $O(n)$ symmetry. It is important to
note that both terms on the right hand side (r.h.s.) of (\ref{H}) may
contain local quadratic terms which may be used to choose them in such
a way that the lattice Fourier transform of ${\epsilon}_{ij}$ behaved as
\begin{equation}
	\epsilon({\bf k})|_{{\bf k}\to0}\propto {\bf k}^2,
	\label{k20}
\end{equation}
which is convenient for implementation of the RG techniques in the case
of homogeneous (ferromagnetic) ordering \cite{wilson} which will be
assumed throughout the paper.

The spin models differ from the general $n$-vector case in that the length
of vectors ${\bf s}_i$ is fixed which for definiteness will be chosen to
be equal to unity \cite{n-vector-models}.  In the functional integral
representation (\ref{Z}) this can be accounted for by the product of
the Dirac's delta-functions as
\begin{equation}
	e^{-H_I}=\prod_i\delta({\bf s}_i^2-1)
	=\lim_{u_4\to\infty}\left({\frac{u_4}{\pi}}\right)^{N/2}
	e^{-u_4\sum_i({\bf s}_i^2-1)^2}
	\label{delta}
\end{equation}
where the second equality shows that this case can also be formally
described by Hamiltonian (\ref{H}) with infinitely strong local quartic
interactions. An important advantage of the layer-cake renormalization
scheme is that such the treatment of such interactions is not more
difficult than of local interactions of finite strength (see section
\ref{partial} below).
\section{\label{func-diff}The functional-differential formalism and the
self-consistency}
A general self-consistent approach to statistical models
within the functional formalism was amply discussed in
\cite{gamma_exp,tokar_new_1997,tan_topologically_2011,arXiv16} so below
only short explanations will be given of the formulas that will be needed
in subsequent calculations.

The Fourier transformed exact (i.e., fully renormalized) CF (\ref{ss_av}) 
may be cast in the form
\begin{equation}
	G^R({\bf k}) = \frac{1}{\epsilon({\bf k})+r^R({\bf k})}
\label{GR(k)}
\end{equation}
where $r^R$ is the exact mass operator. As is seen, (\ref{GR(k)})
expresses one unknown function through another unknown function so a
natural question arises why complicate matters by introducing additional
quantity that is equivalent to the existing one? The answer is that
in many cases the mass operator can be approximated by a constant thus
making its self-consistent calculation simpler because one would need
to solve an equation for a single number instead of a function. The two
cases pertinent to this study are the single-site approximation (SSA)
and the LPA:
\begin{eqnarray}
	\label{ssa}
	r^R_{ij}\simeq r\delta_{ij}\mbox{\ (SSA)}	\\
	\label{lpa}
r^R({\bf k})\simeq r^R({\bf k=0}) =r\mbox{\ (LPA)}.
\end{eqnarray}
As is seen, in both cases the mass operator is site-diagonal
and/or momentum-independent so the approximate CF in both cases has the form
\begin{equation}
G({\bf k}) = \frac{1}{\epsilon({\bf k})+r}.
\label{G(k)}
\end{equation}
The difference lies in the self-consistency conditions which $r$
must satisfy in each case. It is pertinent to note that this form of
renormalized CF implies that the critical exponent $\eta$ is equal to
zero because $r=0$ at the critical point and from (\ref{k20}) it follows
that $G\sim 1/{\bf k}^2$ as $|{\bf k}|\to0$. This deficiency of the LPA
is well known makes the approximation suitable for models in $d\geq3$
\cite{local_potential}. It is for this reason that only 3D systems will
be considered in the explicit calculations below.

Equations for a self-consistent calculation of $r$ can be derived
with the help of the ambiguity in the separation in the quadratic and
the interaction parts in (\ref{H}) mentioned in the previous section
\cite{gamma_exp,tokar_new_1997,tan_topologically_2011,arXiv16}. By adding
and subtracting from the two parts of $H$ on the r.h.s.\ of (\ref{H})
a local quadratic term $r{\bf s}_i\cdot{\bf s}_i/2$ the partition
function (\ref{Z}) may be cast in the functional-differential form 
\cite{hori_approach_1962,vasiliev1998,gamma_exp,tokar_new_1997}
\begin{equation}
	Z[{\bf h}] =
\det(2\pi \hat{G})^{n/2}
	\left.\exp\left(\frac{1}{2}{\partial_{\phi_\sigma}}
\hat{G}{\partial_{\phi_\sigma}}\right)e^{\frac{r}{2}{\boldsymbol\phi}_i
\cdot{\boldsymbol\phi}_i
-H_I[{\boldsymbol\phi}]+{\bf h}_i\cdot{\boldsymbol\phi}_i}
\right|_{\boldsymbol\phi=0}
	\label{Zdiff}
\end{equation}
which can be reduced to (\ref{Z}) with the use of the operator identity
\begin{equation}
\det(2\pi \hat{G})^{n/2}
	\exp\left(\frac{1}{2}{\partial_{\phi_\sigma}}
	\hat{G}{\partial_{\phi_\sigma}}\right)=
	\int D{\bf s}\exp\left(
	-\frac{1}{2}{s}_\sigma \hat{G}^{-1}{s}_\sigma +
{\bf s}_i\cdot\partial_{{\boldsymbol\phi}_i}\right).
	\label{exp_delta}
\end{equation}
To further shorten expressions with summations over the sites
a vector-matrix notation has been introduced in (\ref{Zdiff}) and
(\ref{exp_delta}) with the $N$-component vectors defined as ${\bf s}=[{\bf
s}_i]$ and the $N\times N$ matrix $\hat{G}=||G_{ij}||$ is the propagator
(\ref{G(k)}) in the lattice coordinates.

The source field ${\bf h}$ in (\ref{Zdiff}) can be disentangled
from the action of the second-order differential operator to give
\cite{tokar_new_1997}
\begin{equation}
	Z[{\bf h}]=e^{\frac{1}{2}{h_\sigma}\hat{G}{h_\sigma}}
	e^{{h_\sigma}\hat{G}\partial_{\phi_\sigma}}
	\left.S[{\boldsymbol\phi}] \right|_{\boldsymbol\phi=0}
	=e^{\frac{1}{2}{h_\sigma}\hat{G}{h_\sigma}} S[{\bf h}\hat{G}]
	\label{Z-R}
\end{equation}
where
\begin{equation}
S[{\boldsymbol\phi}]=\det(2\pi \hat{G})^{n/2}
	\exp\left(\frac{1}{2}{\partial_{\phi_\sigma}}
	\hat{G}{\partial_{\phi_\sigma}}\right)
	e^{\frac{r}{2}{\boldsymbol\phi}_i\cdot{\boldsymbol\phi}_i
	-H_I[{\boldsymbol\phi}]}
	\equiv e^{-U^R[{\boldsymbol\phi}]}
\label{R}
\end{equation}
is the generating functional of the S-matrix
\cite{hori_approach_1962,vasiliev1998,gamma_exp,tokar_new_1997} and $U^R$
its connected part.

According to (\ref{ss_av}) the pair CF can be found from (\ref{Z-R}) as
\begin{equation}
	G^R_{\sigma ij}=\left.\frac{\partial^2\ln Z[{\bf h}]}
	{\partial h_{\sigma i}{\partial h_{\sigma j}}}\right|_{\bf h=0}
	=G_{ij}- 
	G_{il}\left.\frac{\partial^2 U^R[{\boldsymbol\phi}]}
	{\partial \phi_{\sigma l}{\partial \phi_{\sigma l^\prime}}}
\right|_{\boldsymbol\phi=0}G_{l^\prime j}
	\label{def-G}
\end{equation}
which establishes a relation between $G^R$ and $G$ that can be used
to derive the self-consistency equation. Though relation (\ref{def-G})
is exact, the simple form of $G$ (\ref{G(k)}) does not make possible 
its identification with $G^R$ by simply setting the second term on the r.h.s.\
to zero. The equality $G^R=G$ can be satisfied only approximately
and the SSA is obtained by requiring that all matrices in (\ref{def-G})
are diagonal.  In this case the second term in (\ref{def-G}) will vanish
provided the second derivative of $U^R$ will be equal to zero
when $l=l^\prime$. This approximation is adequate when the correlations
are strongly localized which means large $r$ or small correlation
length $\xi\propto r^{-1/2}\ll 1$ in the lattice units (l.u.). Explicit
expression for $U^R$ in this case can be found from (\ref{R}) which
factorizes into the product of identical functions of ${\boldsymbol\phi}_i$
\cite{gamma_exp}.  Correspondingly, $U^R$ becomes the sum of single-site
contributions which can be used to derive the SSA equation.

SSA gives a reasonably accurate description
of the first order phase transitions but fails near the critical points
\cite{gamma_exp,tokar_new_1997,tan_topologically_2011} where the LPA 
condition (\ref{lpa}) is more appropriate because it approximates the CF in
the large wavelength limit suitable for the critical region. Formally 
(\ref{lpa}) holds when the Fourier transform of the second derivative of 
$U^R$ in (\ref{def-G}) term is equal to zero: 
\begin{equation}
\left.\frac{\partial^2 U^R[{\boldsymbol\phi}]}
{\partial \phi_{\sigma,{\bf k} }{\partial \phi_{\sigma,{\bf -k}}}}
\right|_{{\boldsymbol\phi=0},{\bf k=0}}=0.
\label{U_2=0}
\end{equation}
where 
\begin{equation}
	{\phi}_{\sigma,{\bf k}}
=N^{-1/2}\sum_je^{-\mathrm{i}j\cdot{\bf k}}{\phi}_{\sigma j}.
	\label{phi_k}
\end{equation}
The effective potential $U^R$ in the long wavelength limit ${\bf k\to0}$
can be calculated with the use of the self-consistent RG equation
derived below.
\section{The renormalization scheme}
In the functional-differential formalism the RG is naturally introduced
through (\ref{R}) cast in the form
\begin{equation}
	e^{-U^R[{\boldsymbol \phi}]}=
	\exp
\left(\frac{1}{2}\sum_{\bf k}\partial_{{\phi}_{\sigma,{\bf k}}}
G({\bf k})\partial_{{\phi}_{\sigma,{\bf-k}}}
\right)e^{-U^0[{\boldsymbol \phi}]}
	\label{URU0}
\end{equation}
where
\begin{equation}
	U^0[{\boldsymbol\phi}] 
	=-\frac{r}{2}{\boldsymbol\phi}_i\cdot{\boldsymbol\phi}_i
	+H_I[{\boldsymbol\phi}]+\mbox{(f.i.t.)}.
	\label{U0}
\end{equation}
Here general notation ``(f.i.t.)''  has been introduced for the
field-independent terms originating from various normalization constants
and may also depend on physical quantities. For example, as follows
from (\ref{R}), in (\ref{U0}) $\mbox{(f.i.t.)}=-\ln\left[\det(2\pi
\hat{G})^{n/2}\right]$.  Such terms are necessary for the calculation of
the absolute value of the free energy but it will not be calculated
in the present paper such terms for brevity will be designated by the
acronym. In case of necessity they can be easily recovered.

The continuous renormalization procedure (discrete renormalization
can be performed similarly) in the functional-differential formalism
proceeds as follows. First the exponentiated operator in (\ref{URU0})
is represented as an integral over some parameter $t$ of commuting
$t$-dependent operators.  For definiteness the parameter can be chosen to
vary from $t=0$ to some maximum value $t_{end}$ with $t=0$ corresponding
to the ``bare'' functional $U^0$ with the fully renormalized $U^R$ being
recovered and at $t=t_{end}$.  Because of the commutativity the integration
can be stopped at any $0\le t\le t_{end}$ which defines a partially
renormalized effective potential $U[{\boldsymbol\phi},t]$. Obviously
that its form will depend on the infinitesimal operators used which
can be defined in an infinite number of ways with different choices
corresponding to different renormalization schemes.  Though $U^R$
at the end of the renormalization should be the same, the RG flow will
strongly depend on the scheme chosen and in approximate calculations may
significantly influence the computational efficiency and the accuracy
of the results obtained.

The main goal of the present paper is to introduce a renormalization
scheme that in the case of lattice models has shown good accuracy in
the calculation of non-universal quantities in 3D $n$-vector models
(see below) and should be suitable for renormalization of models with
very strong local potentials, such as the formally infinitely strong
interactions in the spin models (\ref{delta}). The scheme is a lattice
generalization of the rotationally invariant version of \cite{1984}
which has been achieved via the use of the layer-cake representation
\cite{lieb2001analysis} for the propagator (\ref{G(k)}).
\subsection{The layer-cake representation \cite{lieb2001analysis}}
The representation is defined for non-negative functions so below we will
assume that $G({\bf k})>0$. In this case the layer-cake representation can
be introduced through a simple identity valid for any positive function
\begin{equation}
	G({\bf k})=\int_0^{G({\bf k})}1\,dt^\prime
	=\int_0^\infty \theta[G({\bf k})-t^\prime]dt^\prime.
	\label{layer-cake}
\end{equation}
Substituting this in (\ref{URU0}) we obtain the necessary split of the
operator into infinitesimal parts to be used in the further incremental
renormalization.  As explained above, the partly renormalized effective
interaction $U[{\boldsymbol\phi},t]$ in this case satisfies the equation
\begin{equation}
	e^{-U[{\boldsymbol \phi},t]}=	\exp
\left(\frac{1}{2}\sum_{\bf k}
\int_0^{t} \theta[G({\bf k})-t^\prime]dt^\prime 
\partial_{{\phi}_{\sigma,{\bf k}}}
\partial_{{\phi}_{\sigma,{\bf-k}}}
\right)e^{-U^0[{\boldsymbol \phi}]}
	\label{U(t)}
\end{equation}
or, equivalently, the functional-differential evolution equation
\begin{equation}
	U^\prime_t=\frac{1}{2}\sum_{\bf k}\theta[G({\bf k})-t]
\left(	\partial_{{\phi}_{\sigma,{\bf k}}}
	\partial_{{\phi}_{\sigma,{\bf-k}}}U
	-\partial_{{\phi}_{\sigma,{\bf k}}}U
	\partial_{{\phi}_{\sigma,{\bf-k}}}U\right).
	\label{RG1}
\end{equation}
with the initial condition $U[{\boldsymbol \phi},t=0]=U^0[{\boldsymbol
\phi}]$. Equation (\ref{RG1}) has a typical structure of the
most essential part of the exact RG equations derived, e.g., in
\cite{wilson,nicoll_exact_1976} (see extensive bibliography to more
recent literature in \cite{berges_non-perturbative_2002}) and differs
only in the specific choice of the function $\theta$ and in the absence
of the change of variables usually made to obtain RG equation in the
scaling form.  This change can be easily performed in the case of
necessity but we mostly will not use it because it introduces into the
equation the largest critical exponent $d$ which is completely trivial
by simply reflecting the fact that the free energy grows with the linear
system size $L$ as $L^d$. However, being the largest Lyapunov exponent
of the equation it severely hampers its numerical stability.

\begin{figure}
	\begin{center}
	\includegraphics[viewport = 0 0 325 200]{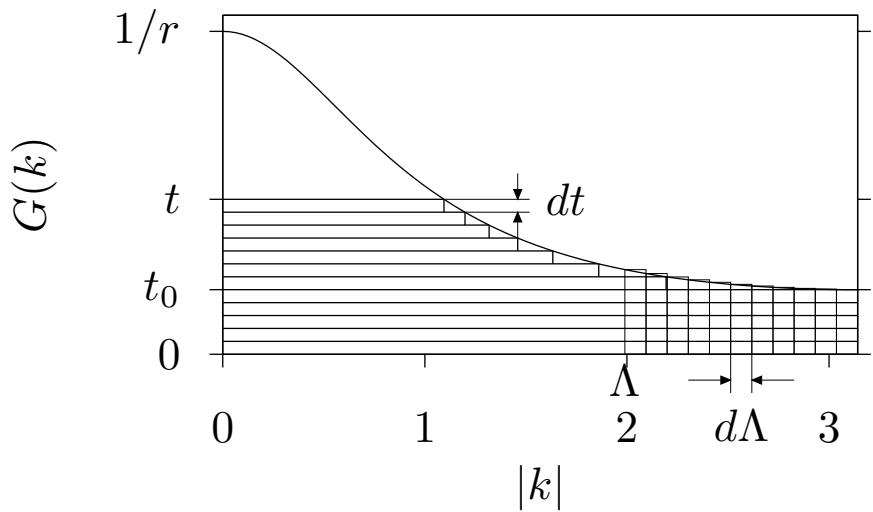}
	\end{center}
	\caption{\label{figure1}Schematic illustration of difference
	between the infinitesimal elimination steps in the layer-cake
	(horizontal slices) and the Wilsonian (vertical slices)
	renormalization schemes shown for 1D propagator (\ref{G(k)}).}
\end{figure}

The difference of the layer-cake renormalization scheme from the
conventional RG procedure is that in (\ref{RG1}) the field variables are
eliminated by infinitesimal layers in the direction orthogonal to that in
the Wilsonian renormalization, as illustrated in figure \ref{figure1}
for 1D case. From this figure and from (\ref{G(k)}) it also can be
seen that the maximum value of the evolution variable $t_{end}=1/r$
in all dimensions.
\subsection{\label{partial}Exact partial renormalization}
To begin with let us solve (\ref{RG1}) in the exactly solvable case
when $\theta=1$ by a method that will be later generalized to the general
case $\theta\not=1$. To this end let us first assume that $U$ remains local
throughout the evolution and express it through the Fourier components
(\ref{phi_k}) as (for simplicity the case $n=1$ is used; for
general $n$ case see \cite{1984})
\begin{equation}
U[{\phi},t] = N\sum_{l=0,\{\bf{k}_m\}}N^{-l/2}u_l(t)
\Delta\left(\sum_{m=1}^l{\bf k}_m\right) \prod_{m=1}^l\phi_{{\bf k}_m} 
	\label{U(k)}
\end{equation}
where $\Delta$'s are the lattice delta-functions.  First we note that
because of the locality the coefficients of the expansion do not depend
on the momenta and so one-to-one correspondence can be established between
(\ref{U(k)}) and the function
\begin{equation}
	u(x,t)=\sum_{l=0}^\infty u_l(t)x^l.
	\label{u(x)}
\end{equation}

The action of the second derivative on the r.h.s.\ of (\ref{RG1}) on
(\ref{U(k)}) amounts to the elimination in each term of two field variables 
which momenta cancel out in the delta-functions argument so the result do
not depend on ${\bf k}$. The two field normalization factors $N^{-1/2}$ 
combine with the sum over the momentum to produce
\begin{equation}
	N^{-1}\sum_{\bf k}1 = 1.
	\label{sum=1}
\end{equation}
The summation over ${\bf k}$ in the second term on the r.h.s.\ of
(\ref{RG1}) is even more trivial because it is lifted by one of
the delta-functions. So the evolution equation in terms of $u(x,t)$
(\ref{u(x)}) reads
\begin{equation}
	u_t = \frac{1}{2}(u_{xx}-u_x^2).
	\label{diff}
\end{equation}
As is easily seen, it is the conventional diffusion equation for 
$e^{-u}$ which solution with the use of the diffusion kernel reads
\begin{equation}
	e^{-u(x,t)}=\frac{1}{\sqrt{2\pi t}}\int dx_0 
	\exp{\left(-\frac{({x-x}_0)^2}{2t}\right)}e^{-u^0(x_0)}
	\label{u0ini}
\end{equation}
where $u^0$ corresponds to (\ref{U0}).

An important property of the layer-cake renormalization scheme is that,
as can be seen from figure \ref{figure1}, when $t$ varies in the interval
from zero to
\begin{equation}
	t_0=\min_{\bf k}G({\bf k})=\frac{1}{\max_{\bf k}\epsilon({\bf k})+r}
	\label{t0}
\end{equation}
the theta function is always equal to unity and does not depend on
${\bf k}$.  This means that (\ref{RG1}) can be integrated from $t=0$ to
$t_0$ exactly.  This property of the RG equation (\ref{RG1}) is especially
useful in the case of the spin models where the delta-function form of
the interaction makes its numerical solution virtually impossible. In
(\ref{u0ini}) the initial probability distribution is smeared by
the Gaussian of width $O(\sqrt{t_0})$ which would lead to weaker
interactions in the renormalized distribution which in some cases may
even admit perturbative treatment. Moreover, if one decides to proceed
with the perturbation theory, there is no need to stop integration at
$t_0$. The general expression (\ref{URU0}) represents a closed form of
the perturbative expansion so by choosing in (\ref{u0ini}) some arbitrary
value $t=g$ and simultaneously subtracting $g$ from $G$ in (\ref{URU0})
one would arrive at the expansion with the renormalized local potential
and the propagator
\begin{equation}
	\tilde{G}({\bf k}) = {G}({\bf k})-g.
	\label{G-g}
\end{equation}
Here $g$ can be chosen to be equal, for example, to the average
value of $G$ within the Brillouin zone (BZ): $g=N^{-1}\sum_{\bf k}G({\bf k})$
\cite{gamma_exp,tokar_new_1997}. The Feynman diagrams in this case
would simplify because the tadpole contributions would be implicitly
accounted for by the redefined propagator (\ref{G-g}).  This is
analogous to the normal ordering in the quantum theory.  Furthermore,
when $r>>\epsilon({\bf k})$ the variation of $G({\bf k})$ within BZ will
be small and its approximation by the average $g$ justifiable. In this
case contributions due to $\tilde{G}\approx0$ may be dropped which amounts
to the SSA \cite{gamma_exp,tokar_new_1997}. Other kinds of perturbative
treatments are possible in this case so the non-linear local couplings
need not be small.  For example, one may resort to the expansion in
real space by exploiting the exponential attenuation of $\tilde{G}$
with distance \cite{gamma_exp}
\begin{equation}
	G_{ij}\sim e^{-|i-j|/\xi}.
	\label{exp_gamma}
\end{equation}
Another possibility is to expand directly in $\tilde{G}({\bf k})$
by estimating the magnitude of the diagrams by the number of lines.
\subsection{RG equation in the LPA}
In the Fourier representation (\ref{U(k)}) the locality means
independence of the expansion coefficients $u_l$ of the momenta ${\bf
k}_m$. The LPA would have been easily justified if the dependence of
$u_l(\{{\bf{k}_m}\},t)$ on the momenta were weak at all $t$. This,
however, is not the case because in the interval $t_0<t\le t_{end}$,
as can be seen from figure \ref{figure1}, the field components near
the BZ boundary are renormalized the least and retain the value
$u_{l>2}(t)\simeq u_l(t_0)$ throughout the whole renormalization
process. But, for example, at the critical point the interactions
are known to attenuate to zero as $t\to\infty$ \cite{wilson} so the
coefficient $u_l(\{\bf{k}_m\},t\to\infty)$ in different regions of the
momentum space will vary in from $0$ to $u_l(t_0)$ and the latter can
be not small. The cause of this is that unlike in the range $0\le t\le
t_0$ where all field components undergo the same renormalization above
$t_0$ the layers shrink from the whole BZ to zero and $\phi_{\bf k}$
with larger ${\bf k}$ are renormalized less. Still, on the basis of
figure \ref{figure1} it is reasonable to assume that at a given $t$
all $\phi_{\bf k}$ with ${\bf k}$ within the closed surface defined
by the condition $\theta[G({\bf k})-t]=1$ are renormalized similarly
during the evolution from $t_0$ to $t$ and so their interactions should
be of similar strength independently of ${\bf k}$. So within this region
the reasoning of the previous section approximately apply with the only
difference that instead of (\ref{sum=1}) the restricted summation over
${\bf k}$ produces the factor
\begin{equation}
	p(t)=\frac{1}{N}\sum_{\bf k}\theta\left[G({\bf k})-t\right].
	\label{p}
\end{equation}
Thus, in the LPA the locality is restricted to the bounded region of field
momenta which shrinks to zero at the end of the renormalization. This,
however, is sufficient for our purposes because in the ferromagnetic
case the source field ${\bf h}$ in (\ref{Z-R}) can be restricted to
the component ${\bf h}_{\bf k=0}$. This precludes the possibility of
studying corrections to $G({\bf k})$ with ${\bf k\not=0}$ but is still
compatible with the {\em ansatz} (\ref{G(k)}).

In the second term on the r.h.s.\ of (\ref{RG1}) the field differentiated
terms from (\ref{U(k)}) are multiplied pairwise and the summation
over ${\bf k}$ becomes trivial due to the delta-functions.  However,
the products of the type $u_lu_m$ are additionally multiplied by the
factor $\theta[G({\bf k})-t]$ with ${\bf k}=\sum_{m=1}^{l-1}{\bf k}_m$
(one of ${\bf k}_m$ disappears due to the summation) which is not equal
to unity for $t>t_0$.  The LPA in this case can be justified under the
assumption that the summation over ${\bf k}_m$ can be restricted to the
field components with sufficiently small momenta when the theta-function
is still equal to unity. Within these approximations the $O(n)$-symmetric
LPA equation reads
\begin{equation}
	u_t = \frac{1}{2}\left[p(t)\nabla_{\bf x}^2 u 
	- (\nabla_{\bf x} u)^2\right].
	\label{LPA}
\end{equation}
It generalizes on the lattice systems the equation derived for the
rotationally symmetric case in \cite{1984}.

Explicit form of function $p(t)$ in (\ref{p}) is easily found as follows. 
Taking into account that the value of 
theta-function does not change when its argument is multiplied by a positive 
quantity, with the use of (\ref{G(k)}) and the fact that $G,t>0$ the integrand 
in (\ref{p}) can be transformed as
\begin{equation}
	\theta[(\epsilon({\bf k})+r)^{-1}-t)]
	=\theta[t^{-1}-r-\epsilon({\bf k})]
	=\int_0^{t^{-1}-r}dE\delta[E-\epsilon({\bf k})].
	\label{identities}
\end{equation}
Substituting this in (\ref{p}) one gets
\begin{equation}
	p(t)=D_{tot}(t^{-1}-r)
	\label{p2}
\end{equation}
where $D_{tot}(E)$ is the integrated density of states (DOS) corresponding
to the quasiparticle band with dispersion $\epsilon({\bf k})$. In the
calculations of section \ref{numerical} $D_{tot}(E)$ was obtained with
the use of the interpolation expressions  derived in \cite{Jelitto1969609}
for DOS of the SC, BCC and FCC lattices. However, because the integrated
DOS $D_{tot}(E)$ is much less structured than the DOS itself it can be
easily obtained by straightforward numerical integration of (\ref{p})
with the use of the Monkhorst-Pack technique \cite{special_points}. In
case of necessity the part $\sim k^2$ describing the critical behaviour
can be subtracted from $\epsilon({\bf k})$ and treated separately
analytically. Thus, in the cases of pair spin interactions of arbitrary
range \cite{ducastelle,Zunger2004} the calculation of $p$ should be
unproblematic.

It is worth noting that the exact partial renormalization is also
described by the RG equation (\ref{LPA}). As is easily seen, when
$t<t_0=[\max\epsilon({\bf k})+r]^{-1}$ the argument of $D_{tot}$ in
(\ref{p2}) exceeds the quasiparticle bandwidth $\max\epsilon({\bf k})$
so the total DOS, hence, $p(t)$ remains equal to unity in this range
and the $O(n)$ generalization of (\ref{diff}) follows. This means, in
particular, that (\ref{LPA}) is valid in the whole range of allowed $t$
values from $t=0$ to $t=t_{end}$.
\section{Gaussian model below $T_c$}
For better understanding of both computational and physical aspects
of the solution of the Ising model in the ordered phase considered
in section \ref{order} it is instructive to first discuss the exactly
solvable Gaussian model at temperatures below $T_c$. The $n=1$ model
corresponds to $H_I$ in (\ref{H}) containing only the local quadratic
term $u^0_2x^2$ in the LPA notation. The coefficient is assumed to be
proportional to $T-T_c$ and is positive above the critical temperature
and negative otherwise.  The functional integral (\ref{Z}) exists
only above $T_c$ so the discussion of the solution  below $T_c$ is
inevitably speculative. Still, the exact solution can be continued into
this region as follows. It is convenient to write the quadratic term
as $u_2^0=(2c_0)^{-1}$ so the solution of the equation reads
\begin{equation} 
	u^{G}(x,t)=\frac{x^2}{2(t+c_0)}+\mbox{(f.i.t.)}.\label{u_G}
\end{equation} 
The linear in $x$ term was neglected because we are interested only
in the symmetric solution. As is seen, when $c_0>0$ the solution is smooth
and well defined at all $t$. But below $T_c$ the initial condition at $t=0$
becomes negative which means $c_0<0$ so the solution (\ref{u_G}) exhibits
singularity at $t=-c_0$. The singularity is non-integrable so the solution
cannot be extended above $t_{end}=-c_0$ where the quadratic term 
diverges $u^G_2\to-\infty$.

Physical meaning of this divergence as well as the above identification
of the value of $t$ at the singularity with the end point of the
renormalization flow can be gained from the consideration the magnetic
susceptibility in the ordered phase (see discussion of this point in RG
context in \cite{maxwell_construction,caillol_non-perturbative_2012}). The
statistical ensemble below $T_c$ in the zero external field consists of
an arbitrary mixture of two pure phases with saturated magnetisations
$\pm m_0$ so in the coexistence region magnetisation may take any value
in the interval $m\in[-m_0,m_0]$. But an infinitesimal external field
$h=0^\pm$ will bring the magnetisation either to $+m_0$ or to $-m_0$
which in the case $m\not=\pm m_0$ means an infinite susceptibility.
However, if the ensemble magnetisation is saturated and is equal, e.g.,
to $m_0>0$ and the infinitesimal field is also positive the susceptibility
will remain finite which means that at $h=0$ it is discontinuous.

This reasoning can be formalized in terms of the LPA solution $u(x,t)$ 
as follows. According to (\ref{s_av}) the magnetisation in the homogeneous 
(ferromagnetic) $n=1$ vector model is
\begin{equation} 
	m = \frac{1}{N}\frac{\partial\ln Z(h)}{\partial h} = x-u_x^R/r
	\label{m} 
\end{equation} 
where $x=h/r$, $u=u(x,t_{end})$ and in the last equality use has been
made of (\ref{Z-R}), (\ref{R}), (\ref{U(k)}) and (\ref{G(k)}). From
(\ref{m}) the susceptibility is found as
\begin{equation}
	\chi = m_h = \frac{1}{r}- \frac{1}{r^2}u_{xx}^R.
	\label{chi}
\end{equation}
At finite temperature the correlation length, hence, $r$ remains finite,
so it is the second derivative in (\ref{chi}) that is responsible the 
infinite value of $\chi$ in agreement with (\ref{u_G}). It seems
that with $u_{xx}\to-\infty$ the self-consistency condition (\ref{U_2=0}) is 
impossible to satisfy. But as mentioned above, there is a jump in the 
susceptibility at $h=0$ and in the ordered state with $m=m_0>0$ it remains
finite as $h\propto x=0^+$ which implies that 
\begin{equation}
u_{xx}|_{x\to0^+}=0 
\label{u_2=0}
\end{equation}
can be used as the self-consistency condition in the LPA both below and
above $T_c$.

In physical models $u_2$ also changes sign at the critical temperature
\cite{wilson}. The boundedness of the Hamiltonian in this case is provided
by higher powers of the field variables, as, e.g., in \ref{ini}. However,
because the differential equation (\ref{LPA}) is local in $x$, the
negative curvature near $x=0$ may drive the solution toward the singular
trajectory (\ref{u_G}) even in the case of bounded Hamiltonians.  This
behaviour has been indeed observed in the attempts to solve (\ref{LPA})
in the ordered region.  In the course of the numerical integration it
proved to be impossible to reach the integration endpoint $t_{end}$ not
only because of the singularity in $t$ but also because as $u_2(t\to
t_{end})\to-\infty$ the approximation of the derivatives by finite
differences became unreliable for any finite step in the field variable.
Because it has been virtually impossible to deal with the singular
functions in the numerical calculations, it has been found necessary
to resort to a $t$-dependent generalisation of the Legendre transform
proposed in \cite{x-yLegendre} (see \ref{x-y}) in order to re-write
the LPA equations in terms of the transformed variables ${\bf y}$ and
the function $v({\bf y},t)$ defined in (\ref{y-x})and (\ref{v-u}). 
In particular, in the case $n=1$ from (\ref{uxx-vyy}) it follows that
\begin{equation}
	\frac{1}{u_{xx}}=\frac{1}{v_{yy}}+t-c
	\label{u_1-v_1}
\end{equation}
which means that the singularity in $u$ (\ref{u_G}) disappears in the
transformed function 
\begin{equation}
	v^G(y,t) = -\frac{y^2}{2(t_{end}-c)}+(f.i.t.)
	\label{v_G}
\end{equation}
provided $t_{end}\not=c$ which can always be satisfied with appropriate
choice of the arbitrary constant $c$.  $v^G$ in (\ref{v_G}) is supplied by
the superscript $G$ because it expectedly is the solution of the Gaussian
model in $y-v$ variables (see equation (\ref{the_eq2}) below). Noticeable
is the fact that the quadratic part of $v^G$ does not depend on $t$ and
thus can be used also at $T_{end}$ as $v_{yy}^R$ for the Gaussian model.

By substituting (\ref{u_1-v_1}) into (\ref{chi}) one finds
\begin{equation}
	\chi^{-1}=r\frac{1+(t_{end}-c)v_{yy}^R}{1-cv_{yy}^R}
	\label{chi-1}
\end{equation}
so that in the Gaussian model below $T_c$ the inverse susceptibility is 
identically zero at all values of $y$ as follows from (\ref{v_G}) and 
(\ref{chi-1}). It can be shown that in the Gaussian model $y$ is simply
proportional to $m$ so the coexistence region comprises all values of $m$
which is natural because formally the saturated magnetisations in the
model are infinite so any $m\in (-\infty,\infty)$ may be found in the
below $T_c$.
\section{\label{numerical}Numerical results}
Though equations in the form (\ref{LPA}) can be readily integrated in the
symmetric phase, they showed lesser stability near the critical point
then the transformed equations, presumably because of the closeness to
the ordering region. Therefore, the transformed equations were used both
above and below $T_c$.
\subsection{The symmetric phase}
In the symmetric phase the transformed equation derived from (\ref{LPAv}),
(\ref{q}) and (\ref{u_xixi}) reads
\begin{equation}
	w_t=\frac{p(t)}{2}\left(\frac{(n-1)w_q}{1+(t-t_0)w_q}
+\frac{w_q+2qw_{qq}}{1+(t-t_0)(w_q +2qw_{qq})}\right)
	\label{RGw}
\end{equation}
where the arbitrary constant $c$ was chosen to be equal to $t_0$. 
With this choice the initial condition is easily found from 
(\ref{y-x}), (\ref{v-u}) and (\ref{q}) as
\begin{equation}
	w(q,t_0)=u(\sqrt{2q},t_0)
	\label{w-ini}
\end{equation}
with $u(x,t_0)$ calculated in \ref{ini}; the superscript ``$(n)$'' was
omitted for consistency with (\ref{RGw}) which holds for all $n\ge1$.  The
self-consistency condition (\ref{u_2=0}) in the fully symmetric case means
that all derivatives $u_{x_\sigma x_{\sigma^\prime}}$ in (\ref{uxx-vyy})
vanish which means that all $v_{y_{\sigma}y_{\sigma^\prime}}$ are
also equal to zero. In the rotationally symmetric case this leads via
(\ref{v_yiyj}) to 
\begin{equation}
	w_q^R|_{q=0}=0.
	\label{scc1}
\end{equation}

Equation (\ref{RGw}) and (\ref{scc1}) with the initial condition
(\ref{w-ini}) has been solved numerically for the $n$-vector spin models
with 
\begin{equation}
	\epsilon_{ij}=-K\delta_{|i-j|/d_{NN},1}+QK\delta_{ij}
	\label{J-NN}
\end{equation}
where $K=|J|/k_BT$, $J$ is the ferromagnetic coupling between 
NN spins, $d_{NN}$ the distance between NN sites and $Q$ is the 
coordination number of the lattice.
The solution has been obtained by the method of lines with the
use of the Fortran LSODE routine \cite{lsode}.  The number of
equations used varied in the range 2-4 thousands until convergence
was reached. The solutions obtained both in the symmetric and
in the ordered phase qualitatively agreed with previous studies
\cite{maxwell_construction,caillol_non-perturbative_2012,arXiv16}.

Near the critical temperature $t_{end}=1/r\to\infty$ but the integration
interval is necessarily bounded so $K_c$ was found by extrapolating
several $r$ calculated at $K\gtrsim K_c$  to $r=0$ according to the
scaling relation
\begin{equation}
	r=C_\pm^{-1}\tau^{2\nu}=C_\pm^{-1}\tau^{\gamma} 
	\label{r(tau)}
\end{equation}
where $\tau = |1-K_c/K|$ and in the LPA where $\eta=0$ $\gamma=2\nu$.
In the Ising model the correlation length both above (+) and below (-)
$T_c$ was estimated on the basis of the asymptotic behaviour in real
space of the Fourier transformed (\ref{G(k)})
\begin{equation}
\xi = \sqrt{K/r}\stackrel{T\to T_c}{\simeq} f_{\pm}\tau^{-\nu}. \label{f+} 
\end{equation}
The critical exponents $\nu$ for consistency were found from the scaled
form of the RG equation (\ref{RGw}) as explained in \cite{1984}. The
values obtained $\nu=0.65$ for $n=1$, 0.71 for $n=2$ and 0.76 for
$n=3$ where similar but larger than those systematized in table 2 in
\cite{berges_non-perturbative_2002} where it can be seen that our values
are closest to those calculated in \cite{berges_non-perturbative_2002}
differing only on 0.01. This apparently is a consequence of a
similar non-perturbative RG approach used by the authors. The
difference can be due to the fact that $\eta$ in the calculations of
\cite{berges_non-perturbative_2002} was not equal to zero but obtained
from the RG equations.

The calculated values of $K_c$ at the critical point found from the LPA
equations (\ref{RGw}) and (\ref{scc1}) are presented in table \ref{T_c}.
\begin{table}
\caption{\label{T_c}Dimensionless inverse critical temperatures of the
$n$-vector spin models on cubic lattices calculated in the LPA. The
errors have been estimated by comparison with precise high temperature
expansions from \cite{n-vector-models} (BCC and SC lattices) and with
the value cited in \cite{FCC_T_c} (FCC lattice).}
\begin{indented}
\item[]\begin{tabular}{@{}llll}
\br
$n$&Lattice&$K_c$&Error\\
\mr
1&FCC&0.1023&0.2\%\\
1&BCC&0.1579&0.3\%\\
1&SC &0.2235&0.8\%\\
2&BCC&0.3225&0.6\%\\
2&SC &0.4597&1.2\%\\
3&BCC&0.4905&0.8\%\\
3&SC &0.7025&1.4\%\\
\br
\end{tabular}
\end{indented}
\end{table}
\subsection{\label{order}Ferromagnetic ordering in the SC Ising model}
To calculate the spontaneous magnetization in zero external field we
still need, according to (\ref{s_av}), to account for the source field $h$. 
The transform \ref{x-y}) in the $n=1$ case is
\begin{eqnarray}
	\label{y-x1}
y=x-(t-t_0)u_{x}\\
	\label{v-u1}
	v=u-\frac{1}{2}(t-t_0)u_{x}^2
\end{eqnarray}
where the arguments of $u(x,t)$ and $v(y,t)$ have been omitted for brevity and
constant $c$ was chosen to be equal to $t_0$ to simplify calculation of the
initial condition (\ref{n=1})
\begin{equation}
	v(y,t_0)=u^{(1)}(x,t_0)|_{x=y}.
	\label{v0=u0}
\end{equation}
The RG equation in this case is 
\begin{equation}
v_t = \frac{p(t)v_{yy}}{2[1+(t-t_0)v_{yy}]}.  \label{the_eq2}
\end{equation}
The dependence of $v^R$ on $h$ at the end of renormalization is defined
parametrically with the use of the expression for $h$
\begin{equation}
x = h/r = y+(1/r-t_0)v_y^R	
	\label{x(y)}
\end{equation}
which follows from (\ref{y-x1}) with $t=t_{end}=1/r$ and from
(\ref{v=u_diff}).  The unknown parameter $r$ in (\ref{x(y)}) is  fixed
by the self-consistency condition (\ref{u_2=0}) which according to
(\ref{uxx-vyy}) and (\ref{u_1-v_1}) reads
\begin{equation}
	v_{yy}|_{h=0^+}=0
\label{v_yy=0}
\end{equation}
where $h$ should be expressed through $y$ according to (\ref{x(y)}).
At $h=0$ there always exists a trivial solution $y=0$ due to the
symmetry. But below $T_c$ more stable solutions with $y\not=0$ appear
which correspond to the states with spontaneous magnetisation
\begin{equation}
	m = y-t_0v^R_y
	\label{m1}
\end{equation}
as can be found from (\ref{m}), (\ref{x(y)}) and (\ref{v=u_diff}).

Numerical solution of equation (\ref{the_eq2}) with the initial condition
(\ref{v0=u0}) and the self-consistency condition (\ref{v_yy=0})
exhibited the same qualitative behaviour as was found previously in
\cite{caillol_non-perturbative_2012}. Namely, the inverse susceptibility
was equal to zero in the interval $(0,m_0)$ (the solution for negative
$m$ is obtained by the symmetry) with a jump to $\chi^{-1}=r$, at $m_0$,
in accordance with (\ref{chi-1}). In particular, it was found that the
Gaussian model solution (\ref{v_G}) describes the field-dependent part of
the solution for the Ising model in the ordered phase so accurately that
the precision of the calculations was insufficient to see the difference.
This can be partly explained by the fact that the $y$-dependent part
of the solution (\ref{v_G}) is stationary and from (\ref{the_eq2})
it can be seen that small deviations from $v^G$ do not grow when $t\to
t_{end}$. However, the singularity at $t=t_{end}$ in the denominator in
(\ref{the_eq2}) is integrable so the deviations should remain finite if
the initial model is not Gaussian.  In view of this the close proximity
of the solution to the Gaussian model seen in the calculations needs 
further investigation.

Similarly, it has been impossible to establish numerically whether the
discontinuity in the inverse susceptibility is genuine or is just a
very steep continuous transition (see discussion of similar behaviour
in \cite{caillol_non-perturbative_2012}).  But anyway the LPA-based
approach is not exact so taking into account the narrowness of the
finite difference step in $y$ of $O(10^{-3})$ within which the inverse
susceptibility changed its value from zero to $r$ and the fact that the
exact behaviour is well understood qualitatively, the values of $m_0$ and
$r$ have been found by interpolation of the solution from the right side
of the jump in $v_{yy}^R$ (see equation (\ref{chi-1})) to the interior
of the interval by assuming that the discontinuity is real. (More details
of the numerical procedure can be found in \cite{arXiv16}.)

\begin{figure}
	\begin{center}
	\includegraphics[viewport = 0 0 283 200]{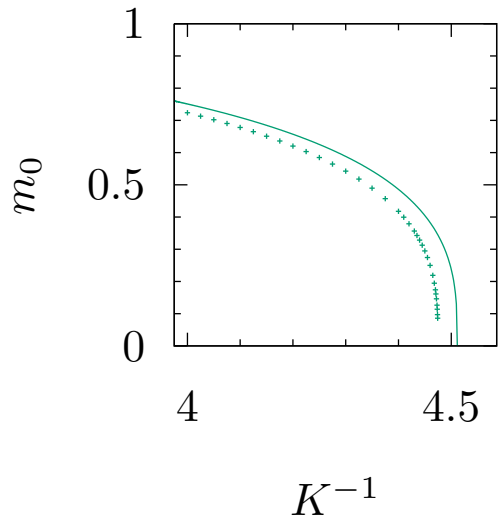}
	\end{center}
	\caption{\label{figure2}Magnetisation curve near the critical
	point as obtained in the LPA (symbols) and the fitting curve
	(\ref{talapov}) to the exact MC simulations with the parameters
	from \cite{talapov_M(t)} (solid line).}
\end{figure}

The magnetisation curve obtained by the above procedure is shown in figure
\ref{figure2} together with the curve
\begin{equation}
m_0(\tau) = \tau^\beta(a_0-a_1\tau^{\theta_W}-a_2\tau)
	\label{talapov}
\end{equation}
that precisely fits the exact MC simulations data \cite{talapov_M(t)}.
In the LPA fit $\beta$ in (\ref{talapov}) was equal to $\nu/2$ and the
Wegner exponent $\theta_W$ approximated by 0.5 \cite{talapov_M(t)}.
In table \ref{amplitudes} the parameters of the fit of the LPA solution
to (\ref{talapov}) is compared with those of \cite{talapov_M(t)}.  As is
seen, parameters $a_{1-2}$ deviate quite appreciably from the (rounded)
exact values of \cite{talapov_M(t)} but it has to be pointed out that
these parameters represent correction terms that enter (\ref{talapov})
multiplied by positive powers of $\tau$. The latter was smaller than 0.1
in the calculations so, on the one hand, the contribution of the terms to
the magnetisation curve was reduced by these factors, on the other hand,
their fit for the same reason was not very accurate and might improve
if more points were calculated.
\begin{table}
\caption{\label{amplitudes}The amplitudes entering the scaling relations
(\ref{r(tau)}), (\ref{f+}) and (\ref{talapov}) calculated in the
present work compared with the high temperature expansions data of
\cite{liu_fisher89} ($C_\pm,f_\pm$) for $\gamma=1.25$ which is closest
to the LPA value 1.3 and the MC simulations of \cite{talapov_M(t)}
($a_0-a_2$).}
\begin{indented}
\item[]\begin{tabular}{@{}llllllll}
\br
Method&$C_+$&$f_+$&$C_-$&$f_-$&$a_0$&$a_1$&$a_2$\\
\mr
LPA&1.06&0.487&0.21&0.22&1.62&0.23&0.37\\
Series and MC &1.06&0.485&0.21&0.25&1.69&0.34&0.43\\
\br
\end{tabular}
\end{indented}
\end{table}

The simulated temperature range was restricted by 10\% distance from
$T_c$ for the following reasons. First, the correlation length at the
lowest temperature was already smaller than the lattice constant (0.8
l.u.) which already is far from the critical region which was of the main
interest in the present study.  Second, at the lowering temperature the
numerical integration considerably slowed down so that convergence to the
self-consistent solution by iterations was difficult to achieve. It is not
excluded, however, that the LPA in the layer-cake renormalization scheme
could describe the low-temperature magnetisation in an asymptotically
exact way. As $T\to0$ $r$ grows very fast so $G({\bf k})$ flattens and
the SSA should be adequate. The SSA does reproduce correctly the leading
asymptotic correction to the saturated magnetisation $m_0=1$ and the
LPA may follow the suite, as was discussed in section \ref{partial}.
However, this can hardly be verified in numerical calculations because
at low temperatures $r$ grows by the Arrhenius law $r\propto e^{AK}$
with $A=O(10)$ so the integration range shrinks as $1/r\sim e^{-AK}$
but at the same time the second derivative $u_{xx}^0$ in (\ref{n=1})
grows as $r^2$ which poses serious computational problems.  By all evidence
the possibility can be reliably verified only analytically.
\section{\label{discussion}Discussion}
The RG equation in the LPA derived in the present paper has made possible
calculation of non-universal quantities in several $n$-vector spin models in
good agreement with the known values obtained within reliable techniques
such as the high-temperature expansions and the Monte Carlo simulations
\cite{n-vector-models,talapov_M(t)}.

The equation, however, has several drawbacks that need be remedied. One
deficiency is that it does not reproduce correctly the critical
exponents. This, however, may be considered as a minor problem because
the exponents are meaningful only asymptotically when the system tends
to the criticality. But in this limit the renormalized Hamiltonian
simplifies and acquires a universal form \cite{wilson}. Thus,
the layer-cake renormalization can be stopped at sufficiently
large value of $t$ and the remaining RG flow accomplished with more
rigorous and accurate perturbative techniques. The necessary Feynman
diagrams with $\tilde{G}$ (\ref{G-g}) can be reduced to the known
expressions by separating $\tilde{G}$ into $G$ and $g$.  The advantage
of this approach is that the necessary parameters of the Hamiltonian
\cite{free_params2RG,free_params2RG2} will be known to a good accuracy
from the LPA solution at $t$. 

But the pure LPA approach can be sufficient when the system is not
too close to the critical point where the knowledge of accurate values
of the exponents is not crucial. For example, from the expression for
magnetisation (\ref{talapov}) it can be found that already 2\% below $T_c$
the error due to the deviation of the LPA $\beta=\nu/2$ from the exact
value introduces the error in $m$ comparable to the error due to neglect
of the correction terms which values depend on non-universal amplitudes
$a_{1-2}$. In this situation magnetisation curve calculated in the LPA
may be more useful from experimental standpoint than the accurate values
of $\beta$ and $\theta_w$ provided by the perturbative RG theory.

Much more serious problem is that in the absence of small parameters the
closed-form approximations to the exact non-perturbative RG equations
are not easy to justify and improve \cite{berges_non-perturbative_2002} so
the good results obtained in several models does not guarantee
that they can be trusted in the cases when the correct answers
obtained within reliable techniques are unavailable.  The LPA is often
substantiated by the derivative expansion of the Hamiltonian which in
the Fourier space corresponds to expansion in powers of the momenta
\cite{berges_non-perturbative_2002,x-yLegendre}.  However, this argument
is valid only in the critical region where the momenta are small. But
if an accurate account of irrelevant variables is needed one has to
deal with the momenta in the whole BZ where the components
of ${\bf k}$ at the boundary reach the values as large as $\pi$ inverse
lattice units which obviously is not a suitable expansion parameter.

However, as was pointed out in the Introduction, in lattice systems
the short-range fluctuations can be efficiently treated within the cluster
approach which may be viewed as a systematic non-perturbative technique
but with poor convergence in the critical region. Still, it can be used
to initialize renormalization by means of the LPA. Schematically this can
be done as follows.  First one assumes that $g$ in (\ref{G-g}) is not
a constant but a function of ${\bf k}$ such that it can be represented
as a finite sum of the lattice Fourier terms
\begin{equation}
	g({\bf k})=\sum_{|l|\leq L^c}g_le^{\mathrm{i}l\cdot{\bf k}}
	\label{g(k)}
\end{equation}
where $L_c$ is a cut-off distance. In the real space the matrix
elements ${g}_{ij}$ will vanish beyond the cluster of radius $L_c$:
$g_{ij}|_{|i-j|>L_c}=0$. Now if the system is far from criticality
the matrix elements of the self-consistent propagator $G$ exponentially
attenuate at large separations (\ref{exp_gamma}) \cite{gamma_exp} and may
be neglected for $|i-j|>L_c$ when $L_c/\xi$ is sufficiently large for the
required accuracy.  In such a case by choosing $g_l\simeq G_{i-j=l}$
in (\ref{g(k)}) one may neglect $\tilde{G}$ in (\ref{G-g}) so the
partition function can be calculated within the cluster generalization
of the SSA with the use of the clusters of radius $\simeq L_c$ (see
\cite{tokar_new_1997,tan_topologically_2011}).

The method just described, however, breaks down in the critical region
which is easily seen at the critical point where $G$ is singular at
${\bf k}=0$.  Because no finite Fourier sum (\ref{g(k)}) can reproduce
the singularity, $\tilde{G}$ in (\ref{G-g}) will be also singular
and so not negligible. Still, one can choose $g_l$ in (\ref{g(k)}) in
such a way that $g({\bf k})$ satisfactorily approximates $G({\bf k})$
throughout the BZ with the exception of a region surrounding
${\bf k}=0$ with $|{\bf k}|<k_c$ where the cut-off $k_c\sim1/L_c$.
Now the contributions due to $g$ can be calculated within the cluster
method while the remaining singular part $\tilde{G}$ for $|{\bf k}|
<k_c$ accounted for within the LPA with the initial local potential
taken from the cluster calculation with the momenta at the vertices set
to zero.  Within this approach the low-${\bf k}$ region will shrink
with growing cluster radius $L_c$ which would validate the gradient
expansion thus justifying the LPA. Besides, diminishing $k_c$ means
smaller renormalized interactions which farther validates LPA which is
known to be exact to the first order in the interactions \cite{1984}.
Finally, as $L_c$ grows the relative contribution of the singular
part into the free energy will diminish which will reduce the errors
introduced by the LPA. Thus, with the use of this hybrid cluster/LPA
approach the results obtained can be validated without resort to the
high-temperature expansions or MC simulations. The feasibility of the
approach is supported by the fact that in the purely cluster approach in
some cases good convergence could be seen with the use of small easily
manageable clusters \cite{tokar_new_1997,tan_topologically_2011}.

In conclusion it is pertinent to note that besides making the LPA-based
approach self-contained, the hybrid cluster/LPA technique would enable
dealing with the short-range cluster interactions that appear, e.g., in
the {\em ab initio} theory of alloys \cite{ducastelle,Zunger2004}. This
opens a possibility of developing an approach which would make possible a
realistic description of the first- and the second-order phase transitions
in lattice systems.
\ack
I would like to express my gratitude to Hugues Dreyss\'e for support
and encouragement.
\appendix\setcounter{section}{0}
\section{\label{ini}Initial condition}
The initial effective interaction at $t_0$ (\ref{u0ini}) for $n=1$ case
straightforwardly generalizes to the $n$-vector spin models (\ref{delta})
with the use of the $n$-dimensional diffusion kernel as
\begin{equation}
	e^{-u^{(n)}({\bf x},t_0)}
=\frac{1}{(2\pi t_0)^{n/2}}\int d{\bf x}_0\,
\exp{\left(-\frac{({\bf x-x}_0)^2}{2t_0}\right)}\delta({\bf x}_0^2-1)
\label{f_ini}
\end{equation}
where the effective interaction at $t=0$ is given by (\ref{delta})
\cite{n-vector-models}.

The $n=1$ case is trivial
\begin{equation}
	u^{(1)}({\bf x},t_0)
	=\frac{x^2}{2t_0}-\ln\cosh \frac{x}{t_0} + \mbox{(f.i.t.)}
\label{n=1}
\end{equation}
where $x=|{\bf x}|$ and (f.i.t.) stands for $x$-independent terms. 

For $n>1$ in the $O(n)$ symmetric case the integral in
(\ref{f_ini}) is convenient to calculated in hyperspherical
coordinates. Choosing the direction of ${\bf x}$ along the first axis
${\bf x}=(x\cos\theta,0,0,\dots,0)$ one gets \cite{n-sphere}
\begin{equation}
	e^{-u^{(n)}({\bf x},t_0)}
{\;\propto\;}e^{-\frac{x^2}{2t_0}}\int_0^{\pi}
e^{\frac{x}{t_0}\cos\theta}\sin^{n-2}\theta\,d\theta.
\label{theta}
\end{equation}
The cases $n=2$ and $n=3$ are given by
\begin{eqnarray}
u^{(2)}({\bf x},t_0)
=\frac{x^2}{2t_0}-\ln I_0\left(\frac{x}{t_0}\right) 
+ \mbox{(f.i.t.)}\nonumber\\
u^{(3)}({\bf x},t_0)
=\frac{x^2}{2t_0}-\ln \left(\frac{t_0}{x}\sinh\frac{x}{t_0}\right) 
+ \mbox{(f.i.t.)}.
	\label{n=2,3}
\end{eqnarray}
where $I_0$ is the modified Bessel function of the first kind.

Explicit expressions for $n>3$ will not be considered in the present paper
so we only note that at large $n>8$ a three-term recurrence relation for
integrals in (\ref{theta}) can be established so additionally only $n=4-8$ 
integrals will need to be calculated explicitly in the large-$n$ case.
\section{\label{x-y}Transformed RG equation}
In our case the Legendre transform suggested in \cite{x-yLegendre} (see
also \cite{local_potential}) should be slightly modified as
\begin{eqnarray}
	\label{y-x}
y_\sigma({\bf x},t)=x_\sigma-(t-c)u_{x_\sigma}({\bf x},t)\\
	\label{v-u}
	v({\bf y},t)=u({\bf x},t)-\frac{1}{2}(t-c)\sum_\sigma u_{x_\sigma}^2({\bf x},t)
\end{eqnarray}
where $\sigma=1,\dots,n$ and $c$ is an arbitrary constant. Here the
independent variables are ${\bf x}$ and $t$ and the effective interaction
is $u$; our aim is to use these relations to re-write (\ref{LPA}) in terms
of ${\bf y}$ and $t$ for function $v$. To this end we first differentiate
Eqs.\ (\ref{y-x}) and (\ref{v-u}) with respect to $x_{\sigma^\prime}$:
\begin{eqnarray}
	\label{y-x_diff}
\frac{\partial y_\sigma }{\partial x_{\sigma^\prime}}=
\delta_{\sigma\sigma^\prime}-(t-c)u_{x_\sigma x_{\sigma^\prime}}\\
	\label{v-u_diff}
 v_{y_\sigma }\frac{\partial y_\sigma }{\partial x_{\sigma^\prime}}
 =u_{x_{\sigma^\prime}} -(t-c)u_{x_\sigma }u_{x_\sigma x_{\sigma^\prime}}.
\end{eqnarray}
(we remind the summation over repeated indices convention).  Substituting
(\ref{y-x_diff}) into (\ref{v-u_diff}) one gets after some rearrangement
a linear system
\begin{equation}
	[\delta_{\sigma {\sigma^\prime}}-(t-c)u_{x_\sigma x_{\sigma^\prime}}]
	(v_{y_{\sigma^\prime}}-u_{x_{\sigma^\prime}})=0.
	\label{Mv-u}
\end{equation}
Because the matrix in this equation in general is not singular it
follows that for all $\sigma=1,\cdots,n$
\begin{equation}
	v_{y_\sigma }=u_{x_\sigma }.
	\label{v=u_diff}
\end{equation}
Differentiation of this with respect to $x_{\sigma^\prime}$ and using
(\ref{y-x_diff}) gives
\begin{equation}
	[\delta_{\sigma\kappa}+(t-c)v_{y_{\sigma}y_\kappa}]
	u_{x_\kappa x_{\sigma^\prime}}
	=v_{y_{\sigma}y_{\sigma^\prime}}.
	\label{uxx-vyy}
\end{equation}
As is seen, for fixed $\sigma^\prime$ one gets a linear system of size
$n$ for expressing $n$ derivatives $u_{x_\sigma x_{\sigma^\prime}}$ in terms
of $v_{y_\kappa y_{\kappa^\prime}}$.

Finally, differentiating Eqs.\ (\ref{y-x}) and (\ref{v-u}) by $t$ and
using Eq.\ (\ref{v=u_diff}) one gets (note the difference with equation
(15) in \cite{x-yLegendre} where the second term on the l.h.s.\ is absent)
\begin{equation}
	u_t+\frac{1}{2}\sum_{\sigma}u_{x_{\sigma}}^2=v_t
	\label{ut-vt}
\end{equation}
so that (\ref{LPA}) can be written as
\begin{equation}
	v_t=\frac{1}{2}p(t)u_{x_\sigma x_\sigma}
	=\frac{1}{2}p(t)\nabla_{\bf x}^2 u
	\label{LPAv}
\end{equation}
where the r.h.s.\ should be expressed in terms of $v_{y_\sigma
y_{\sigma^\prime}}$ with the use of (\ref{uxx-vyy}). Because
(\ref{uxx-vyy}) depends only on the second order derivatives the annoying
negative quadratic terms disappears from (\ref{LPAv}).
\subsection{Fully $O(n)$ symmetric case} 
In the case of full rotational symmetry (\ref{uxx-vyy}) can be solved 
explicitly as follows. Assuming 
\begin{equation}
	v({\bf y},t)=w(q,t)\mbox{\ where\ }q={\bf y}^2/2
\label{q}
\end{equation}
one finds
\begin{equation}
v_{y_{\sigma^\prime} y_\sigma }=\delta_{\sigma^\prime\sigma}w_q
+y_{\sigma^\prime} y_\sigma w_{qq}.
\label{v_yiyj}
\end{equation}
Now denoting the matrix in (\ref{uxx-vyy}) as $\hat{M}$ with the use of 
(\ref{v_yiyj}) one gets
\begin{equation}
M_{\sigma\kappa}=\delta_{\sigma\kappa}+(t-c)v_{y_\sigma y_{\kappa}}
=[1+(t-c)w_q]\left[\delta_{\sigma\kappa}
+y_\sigma y_{\kappa}\frac{(t-c)w_{qq}}{1+(t-c)w_q}\right]
\label{M}
\end{equation}
Substituting this in (\ref{uxx-vyy}) and solving for 
$u_{x_\sigma x_{\sigma^\prime}}$ one arrives at the expressions that eliminates 
$u$ and ${\bf x}$ on the r.h.s.\ of (\ref{LPAv}) as
\begin{equation}
	\nabla^2_{\bf x} u =\frac{(n-1)w_q}{1+(t-c)w_q}
+\frac{w_q+2qw_{qq}}{1+(t-c)(w_q +2qw_{qq})}.
	\label{u_xixi}
\end{equation}
In the case $n=1$ the introduction of $q$ may be superfluous because
with $v_{yy}=w_q +2qw_{qq}$ the equation simplifies \cite{x-yLegendre}.
\bibliographystyle{iopart-num}
\providecommand{\newblock}{}

\end{document}